\documentclass[aps,prl,reprint,superscriptaddress]{revtex4-1}

\usepackage{amsmath}
\usepackage{graphicx}
\usepackage{natbib}
\usepackage{xcolor}
\usepackage{hyperref}

\begin{document}


\title{Glassy dynamics and memory effects in an intrinsically disordered protein construct} 



\author{Ian L. Morgan}
\affiliation{BMSE Program, University of California, Santa Barbara, California 93106, USA}
\author{Ram Avinery}
\author{Gil Rahamim}
\author{Roy Beck}
\affiliation{The Raymond and Beverly Sackler School of Physics and Astronomy \& The Center for Nanoscience and Nanotechnology, Tel Aviv University, Tel Aviv 69978, Israel}
\author{Omar A. Saleh}
\affiliation{BMSE Program, University of California, Santa Barbara, California 93106, USA}
\affiliation{Materials Department, University of California, Santa Barbara, California 93106, USA}


\date{\today}

\begin{abstract}
Glassy, nonexponential relaxations in globular proteins are typically attributed to conformational behaviors that are missing from intrinsically disordered proteins. Yet, we show that single molecules of a disordered-protein construct display two signatures of glassy dynamics, logarithmic relaxations and a Kovacs memory effect, in response to changes in applied tension. We attribute this to the presence of multiple independent local structures in the chain, which we corroborate with a model that correctly predicts the force-dependence of the relaxation. The mechanism established here likely applies to other disordered proteins.
\end{abstract}

\maketitle 



The conformational changes of globular, folded proteins can exhibit glass-like kinetics, typically measured as nonexponential relaxations \cite{Frauenfelder1991,Morozova-Roche1999,Sabelko1999,Brujic2006,Hinczewski2016}.
This behavior is associated with the roughness of the conformational energy landscape, i.e.,~the presence of multiple local free energy minima that are separated by appreciable activation barriers \cite{Onuchic1997}.
Based on studies of random-sequence biopolymers, the heights of the barriers are usually related to either the difficulty in rearranging connected residues within the dense protein core (`topological frustration'), or to kinetic trapping by nonnative contacts (`energetic frustration'), as enabled by the nonspecific nature of the dominant hydrophobic interactions \cite{Ferreiro2014}.

Unlike globular proteins, intrinsically disordered protein regions (IDRs) exhibit a high degree of conformational freedom in their native state \cite{Uversky2002}. IDRs generally are enriched in hydrophilic and charged residues \cite{Uversky2002}, permitting them to assume a dynamic ensemble of structures analogous to those of a random-walk polymer in good solvent, though with the typical addition of some secondary-structure formation \cite{Marsh2010} or other forms of weak attraction between residues \cite{Kornreich2015, Kornreich2016, Malka-Gibor2017, Muller-Spath2010}.
 These structures and interactions are thought to give IDRs a rough,  flat, energy landscape \cite{Solanki2014,Uversky2019}. However, because IDRs have fewer hydrophobic residues than globular proteins,  and no dense core, the barriers on this landscape are small; thus, IDRs would not be expected to exhibit significant frustration, nor, in turn, glass-like kinetics.
\begin{figure}[b!]
\centering
\includegraphics[scale=0.97]{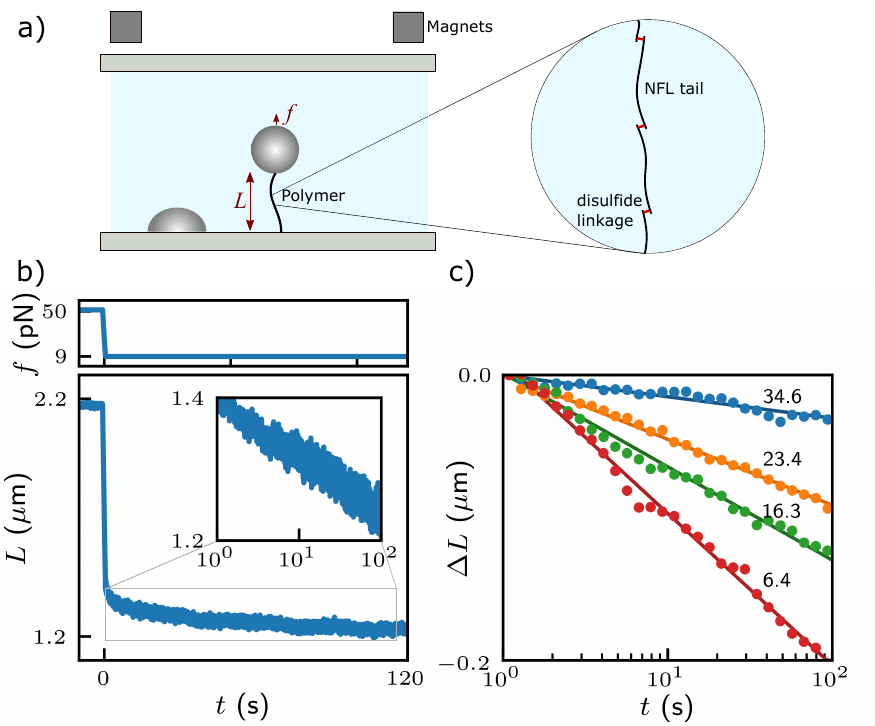}
\caption{(a) Experimental setup: A polymer consisting of multiple NFL IDRs joined by disulfide bonds is stretched with a force $f$ while its extension $L$ is tracked. Stuck beads are tracked to remove drift. (b) Example force-quench experiment on a single polymer: at $t=0$, the force was decreased from $f_1=50$ to $f_2=9$ pN, resulting in a rapid elastic response followed by a slow logarithmic relaxation (inset). (c) 
Typical force dependence of the logarithmic relaxation after quench from $f_1 = 50$ pN, plotted as the compaction, $\Delta L \equiv L(t)-L(t_0)$, after a reference time $t_0=1$ s. At higher $f_2$ (labeled, in pN), the relaxation slows due to hindering of chain shortening by tension. Data points and error bars are the mean and standard error of the mean after logarithmic binning in time (error bars are smaller than points); lines are best-fits to $ b\log \left(t/t_0\right)$.}
\label{Fig1}
\end{figure}

Yet, here we show that single molecules of a model disordered protein construct, consisting of multiple repeats of the intrinsically disordered neurofilament-low (NFL) protein tail region \cite{Pregent2015}, exhibit glass-like behavior in the form of slow, logarithmic relaxations in response to a one-step change in applied tension. Further, when subject to a two-step force-change protocol, the construct displays a nonmonotonic change in the polymers’ extension, a glassy memory effect termed the `Kovacs hump' \cite{Kovacs1964,Bertin2003}. Based on work on other glassy systems \cite{Amir2012, Lahini2017}, we attribute these behaviors to the existence of multiple, independent local structure-forming processes with widely-varying dynamics. We corroborate this picture by showing that the force-dependence of the logarithmic relaxation is well-described by a model that couples an ensemble of parallel structuring processes to Bell-Zhurkov mechanochemistry \cite{Bell1978,Zhurkov1965}. Overall, this work demonstrates glassy behavior in this IDR construct is due to a heterogeneous, distributed mechanism different from the frustration-based ones of certain globular proteins. 

\textit{Methods} - IDR purification and polymer synthesis are described in detail in the supplement \cite{supplement}. In short, single NFL IDRs, each containing 168 amino acids, were modified to carry cysteine residues at each terminus, recombinantly expressed, and purified. These IDRs were polymerized together to form a linear polymer by inducing disulfide bonds between the cysteines. The polymers were terminally labeled with azide and biotin, respectively, allowing specific attachment between a functionalized glass surface and a 2.8-$\mu$m-diameter magnetic bead; this enabled stretching experiments [Fig.~\ref{Fig1}(a)]. 

Experiments were carried out with a custom-built magnetic tweezer setup \cite{Ribeck2008,Innes-Gold2019}, at $T = 20~^\circ$C, in a pH 7 buffer containing 20 mM 2-(N-morpholino)ethanesulfonic acid (MES), 10 mM NaCl, and 0.05\% Tween-20. The stretching force was set by adjusting the distance between a pair of movable magnets and the flow cell surface [Fig.~\ref{Fig1}(a)]. The polymers' end-to-end extension was measured by analyzing the image of the bead \cite{Ribeck2008,Gosse2002}, as captured by a CMOS camera operating at 400 Hz. Instrumental drift was eliminated by simultaneously tracking reference beads stuck to the glass surface, and subtracting their height from that of the experimental beads [Fig.~\ref{Fig1}(a)]; the success of this procedure was demonstrated through control measurements of DNA tethers (data shown in the supplement \cite{supplement}). The stretching force was estimated by analyzing lateral bead fluctuations \cite{Lansdorp2012}, with a typical uncertainty of $\lesssim5 \%$.

For each polymer, the number of monomer tails, $N$, was estimated from the polymer ($L_p$) and monomer ($L_m\approx64$ nm) contour length, $N=L_p/L_m\approx$ 2--29, by assuming a contour length per amino acid of $0.38$ nm and 168 amino acids per monomer. The polymers' contour lengths were estimated from their high force ($\gtrsim 50$ pN) extension. After accounting for $N$, the polymers' polydispersity did not affect our measurements.

During force-quench experiments, the force was changed from $f_1$ to $f_2<f_1$ by moving the magnets away from the flow cell surface. The motion of the magnets lasted $\approx 0.25$ s; we only analyzed extension changes that occur after that, particularly setting the zero of time, $t=0$, as the point at which magnet motion stops. During the motion, the polymer extension changed rapidly due to its entropic elasticity [Fig.~\ref{Fig1}(b)]. The time scale of elastic relaxation is expected to be $\approx 10$ ms, as judged by estimating either the Rouse time \cite{Rouse1953} of the polymer or the relaxation time associated with the drag of the bead; thus, elastic relaxation is unrelated to the observed long timescale extension changes. The relaxation data and analysis code have been made available in a public repository \cite{repository}.
	
\textit{Logarithmic relaxation} - Following a force quench, the polymer extension, $L$, decreased logarithmically in time [Fig.~\ref{Fig1}(b), inset]. As shown in the supplement \cite{supplement}, logarithmic relaxations have been observed to last for up to 3 decades in time, although our analysis focused on 2-decade relaxations [Fig.~\ref{Fig1}(c)]. During relaxation, the extension change was smooth without any detectable discrete transitions of 10 nm or larger, suggesting that the underlying individual compaction events each contribute a length change of order 1 nm.  

We studied the force-dependence of the relaxation by keeping $f_1$ constant, typically 50--70 pN, and quenching to different values of $f_2$.  For all polymers, the relaxation was logarithmic for all $f_2$, with the rate becoming faster for smaller $f_2$ [Fig.~\ref{Fig1}(c)], as expected for a structure-forming process that is hindered by an opposing force \cite{Bell1978,Zhurkov1965}. All relaxations were well-fit by the relation $L(t)= b\log\left(t/t_0\right)$, with $t_0$ being an arbitrary reference point chosen throughout to be $t_0=1$ s, and $b$ corresponding to the log-slope of the relaxation.
	
\begin{figure}[t]
\centering
\includegraphics[scale=0.96]{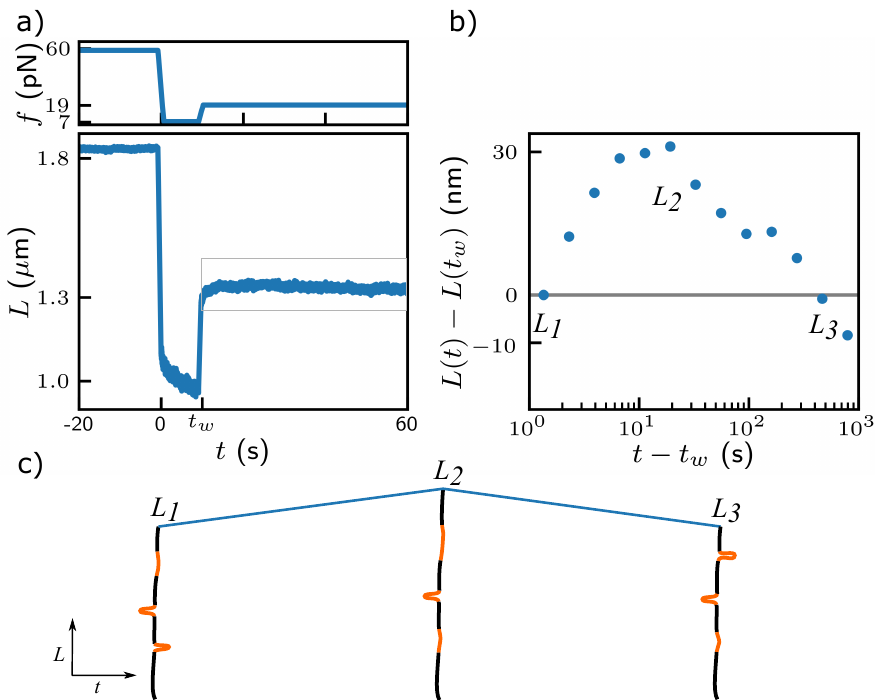}
\caption{(a) Typical two-step experiment: The force was initially $f_1 = 60$ pN, then held at $f_2 = 7$ pN for $t_w = 10$ s, then increased to $f_3 = 19$ pN. (b) Detail of extension dynamics at $f_3$ from the data in (a), showing a nonmonotonic change in $L$, i.e. a Kovacs hump \cite{Bertin2003}. Data points and error bars are the mean and standard error of the mean after logarithmic binning in time (error bars are smaller than points). (c) Cartoon of heterogeneous dynamics within a single IDR domain that result in the Kovacs hump: Incubation at $f_2$ for $t_w$ (left) allows folding of fast segments, but is not long enough to allow folding of slow segments. Application of the higher force $f_3$ causes unfolding of some fast segments, leading to the increase, $L_2>L_1$. At long times, the slow segments finally fold, causing the slow decrease, $L_3<L_2$.}
\label{Fig2}
\end{figure}

\textit{Heterogeneity} - Recent work has suggested there are multiple classes of logarithmically-relaxing systems \cite{Amir2012,Lahini2017}. Among these, Amir \textit{et al.} \cite{Amir2012} identified a mechanism in which the system is highly heterogeneous, consisting of multiple modes that relax independently and with a broad spectrum of timescales. Lahini \textit{et al.} \cite{Lahini2017} showed that this mode structure is associated with an experimentally-observable memory effect, the Kovacs hump \cite{Kovacs1964,Bertin2003}, which here corresponds to the prediction of a nonmonotonic change in extension with time (i.e. an increase followed by a decrease in $L$) after a particular two-step pattern of changes in applied force. Such a behavior unambiguously demonstrates heterogeneity: as the trajectory progresses, certain values of $L$ are reached twice, but are followed by different system behaviors (i.e.~lengthening or contraction). Thus, knowing the single parameter $L$ is not sufficient to predict future behavior. Instead, predicting the correct behavior requires the knowledge of more parameters, i.e.~the status of the diverse modes, which store the system's memory of past force application \cite{Lahini2017,Bertin2003,Mossa2004}.

To assess whether the present IDR system belongs to this `Kovacs class', we tested for the Kovacs hump by subjecting the chain to three successive forces, $f_1, f_2,f_3$,  such that $f_2$ is held for a time $t_w$, and the final force lies between the prior two, $f_1>f_3>f_2$ [Fig.~\ref{Fig2}(a)]. 
We found that single NFL IDR polymers, and not control DNA molecules \cite{supplement}, consistently showed a clear Kovacs hump at $f_3$ ([Fig.~\ref{Fig2}(b)]; observations on 8 other polymers are shown in the supplement \cite{supplement}).  The hump consisted of a slow increase in $L$ followed by a slow decrease, and thus was not related to the fast elastic response of the polymer. We conclude that the polyIDR can be assigned to the heterogeneous, Kovacs-class of aging systems.

\textit{Microscopic view} - The results of Fig.~\ref{Fig2} indicate that the IDR polymers contain multiple independent relaxation modes, but does not clarify their microscopic identity. We posit that each mode corresponds to a different segment of the chain, with each segment able to independently transition from an extended coil to a compact structure. As discussed below, the different segments do not correspond to individual IDRs within the polyIDR chain, but rather to different clusters of residues within each IDR. Indeed, previous work has observed short-range structure, such as salt bridges and residual secondary-structure elements, in the NFL tail domain \cite{Kornreich2015, Kornreich2016, Malka-Gibor2017}.

Such local structure implies the Kovacs hump occurs in the following manner (see also Fig.~\ref{Fig2}(c)): Incubation at the high force $f_1$ converts all segments to the extended coil state. After quenching to $f_2$ and holding for a time $t_w$, a fraction of the segments become structured (i.e.~those with relatively fast dynamics), while the slower segments remain unstructured. Jumping to the final force $f_3$ causes a transition back to the extended state for some fast segments (leading to the initial increase). After a long time, the slow segments become structured (leading to the long-term decrease).

Using the established effect of force on transition kinetics, we can develop a model that quantifies this microscopic picture, and tests it through comparison to data. We focus on the single-step force quench, and assume each IDR consists of $n$ independent segments (and thus that the entire polymer contains $Nn$ such segments). We then adopt the mathematical framework of Amir \textit{et al.} \cite{Amir2012}, and take each segment to relax, on average, exponentially after the force quench, so the contraction dynamics of the $j^{th}$ segment follows $L_j(t) = \alpha(f) e^{-t/\tau_j(f)}$, where $j = 1,2,\ldots,n$. Both $\tau_j$, the contraction timescale, and $\alpha$, the relaxation amplitude, carry a force-dependence. Each structuring event within a single IDR is discrete and stochastic; however, the presence of multiple IDRs in the polymer ($N \gg 1$) means that the measured extension change will follow the exponential ensemble-average behavior.

The force-dependence of $\alpha$ accounts for segment elasticity, which is likely dominated by the flexible coil state. Thus, we take $\alpha(f) =\ell\alpha_0(f)$, where $\ell$ is the coil contour length, and $\alpha_0(f)$ is the relative extension, given by the worm-like chain model \cite{Marko1995} with $l_p = 0.8$ nm, as appropriate for polypeptides \cite{Rief1999}.
Using more nuanced elastic models does not greatly affect our results, as described in the supplement \cite{supplement}.
While different segments likely have different $\ell$, we expect that variation to be small ($<10\times$) compared to the range of values of $\tau_j$ that must underlie the multi-decade dynamics, and thus take $\ell$ to be the same for all segments. 

This mode structure predicts logarithmic relaxation if the characteristic relaxation times are distributed as $P(\tau) \propto 1/\tau$ \cite{Amir2012}. We implement this here by noting it implies a uniform distribution of log-relaxation-time, $P\left(\log\left(\tau\right)\right) \equiv \eta$, where $\eta$ is the density of mode states in log-time units. This means that the log of the time between successive relaxation events is, on average, $1/\eta$. Thus, during a time interval $t$, the number of relaxation events that occurs in a single IDR is $\eta \log t$. Since each event contributes the same contraction, this leads to a logarithmic relaxation. 

To compare to force-quench data, we enforce the condition that there is no relaxation if $f_2=f_1$, reference the relaxation to the time $t_0$, and scale by $N$ to get the extension change of the entire chain: 
\begin{equation}\label{eq1}
L(t,f_2)-L(t_0,f_2) = -N\alpha(f_2) \left[\eta (f_2)-\eta(f_1)\right] \log\left(t/t_0\right)
\end{equation}
As noted, $\eta$ must carry a force dependence, since the transition times  $\tau$ vary with $f$. 

The dependence on force follows from enforcing an Arrhenius dependence of $\tau$ on activation barrier, $\tau = \tau_0 \exp(\Delta G/k_BT)$, and using the Bell-Zhurkov expectation that $\Delta G$ varies as $f\Delta x$, for activation distance $\Delta x$  \cite{Bell1978,Zhurkov1965}. The constraint, $P(\tau) \propto 1/\tau$, corresponds to a uniform distribution of $\Delta G$ \cite{Amir2012}; thus we take 
\begin{equation}
\Delta G_j = G_0 + j f_2 \delta x
\label{eq2}
\end{equation}
where $G_0$ is the barrier at zero force, and $j\delta x$ is the distance between the initial (extended) state and the $j^{th}$ activation barrier. Eq.~\ref{eq2} implies that $P(\Delta G) = 1/f\delta x$, and thus that $\eta =  k_BT/f\delta x$, which when combined with Eq.~\ref{eq1} results in a prediction for the log-slope $b$. 

This analysis indicates useful normalized parameters for the slope and force-quench magnitude are $\bar b \equiv bf_1/Nk_BT\alpha_0(f_2)$ and $\bar f\equiv f_2/f_1$. Indeed, plotting $\bar b$ vs.~$\bar f$ collapses the data [Fig.~\ref{Fig3}], including removing the effects of polydispersity in length, as shown in the supplement \cite{supplement}. The model specifically predicts:
\begin{equation}
\bar{b} = \frac{1}{\rho}\left[1-\frac{1}{\bar{f}}\right]
\label{eq3}
\end{equation}
The single unknown parameter $\rho$ represents the spacing between activation barriers relative to the coil contour length, $\rho \equiv \delta x/\ell$. For our model to be self-consistent, we expect $n\delta x < \ell$ and $\rho < 1/n < 1$.

\begin{figure}[t]
\centering
\includegraphics[scale=1]{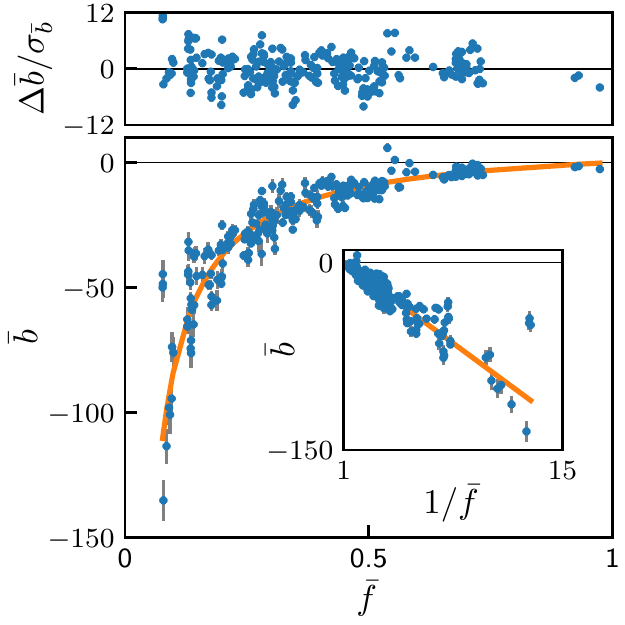}

\caption{Bottom: Dependence of the normalized logarithmic relaxation rate, $\bar b$, on the force-quench magnitude, $\bar f \equiv f_2/f_1$. Each data point represents a single force-quench on a single polymer, with error estimated from the uncertainty in the measured slope. 248 data points, from 16 separate polymers, are shown. The line is a fit to Eq.~\ref{eq3} with best-fit parameter  $\rho = 0.108 \pm 0.004$ (error estimated from bootstrapping,  as described in the supplement\cite{supplement}). Top: standardized residuals of the fit, $\Delta \bar b/\sigma_{\bar b}$, where $\Delta \bar{b}$ is the difference between the data and the fit, and $\sigma_{ \bar{b}}$ is the error estimate of the data.  Inset: The data is linearized by plotting $\bar b$ vs.~ $1/\bar f$.}
\label{Fig3}
\end{figure}

The microscopic model, Eq.~\ref{eq3}, successfully describes the force-quench data, corroborating the picture of multiple independent structured segments that each follow Bell-Zhurkov mechanochemistry. We show this by fitting Eq.~\ref{eq3}, using the single fitting parameter, $\rho$, to the results of 248 different force-quench experiments [Fig.~\ref{Fig3}]. The best-fit is found with $\rho = 0.108 \pm 0.004$. The fit gives a reasonable reduced-chi-squared fitting metric, $\bar{\chi}^2 = 10.7$. Note that $\bar{\chi}^2 \approx 1$ is expected for a statistically valid fit given the stochastic errors of the data; the elevated value here is due either to unknown systematic errors, or to physical effects ignored in our approximate model. The standardized residuals show no systematic deviation [Fig.~\ref{Fig3}, top], strongly suggesting our model captures the key features of the system. 

The best fit estimate of $\rho<1$ is consistent with the physical restriction on the activation barrier spacing, $\delta x < \ell$. The fit value of $\rho$ implies that the upper limit on the number of structured segments per IDR is $n_{max} = 1/\rho \approx 9$.  This value is also consistent with the data: A typical polymer has $N=25$ IDRs; taking $n=9$ structures per IDR, and with each structure contributing a compaction $\alpha \approx 1$ nm, we can estimate the total length change during relaxation, $\Delta L \approx 200$ nm. This is indeed the magnitude of the total length change seen in the force-quench experiments [Fig.~\ref{Fig1}(c)], further supporting our picture. 

The polymeric nature of the construct allows for the possibility of inter-tail interactions \cite{Yu2008,Neupane2014}, however the measured relaxation is likely dominated by multiple intra-tail structures. Given that there are no observable discrete length changes [Fig.~\ref{Fig1}(b)] and the total length change is a small fraction ($<10\%$) of the polymer's contour length, the structure-forming interactions must be quite short-range. With only $N\approx 25$ abutting neighbors, nearest-neighbor inter-tail interactions, contributing $\approx 1$ nm length changes each, could not account for the observed 200 nm length changes.

While the precise identity of the structures is as yet unclear, we can roughly estimate their free energy based on their ability to compact against a known load. Extrapolated to zero force, we find that single-segment structure stability is likely between $3$ and $9 k_BT$; the wide range is due to sensitivity to the choice of $\ell$ \cite{supplement}. This range is reasonable and suggests some possible mechanisms; it encompasses prior estimates of the stability of local structures in IDRs \cite{Solanki2014}, as well as estimates of attractive electrostatic interactions in the NFL tail \cite{Beck2010,Kornreich2015,Kornreich2016,Malka-Gibor2017}.

In summary, our analysis indicates that the NFL IDR has multiple independent structures with a broad distribution of relaxation times. The distribution of relaxation timescales produced a logarithmic relaxation of polymer extension that can last for three decades in time \cite{supplement}. We attribute the long timescales to the slowing of individual compactions by applied force, in analogy to the slowing caused by low temperatures in observations of nonexponential relaxations in globular proteins \cite{Frauenfelder1991}. The heterogeneous, independent nature of the structures was confirmed by the observation of the Kovacs hump. Finally, our picture of IDR compaction dynamics is confirmed by a model that combines Bell-Zhurkov mechanochemistry with a specific distribution of independent segment relaxation times; this model successfully describes the dependence of relaxation rate with force, and produces consistent estimates of microscopic parameters. 

The nonexponential relaxations reported here are similar to those observed in prior work on globular proteins \cite{Frauenfelder1991,Brujic2006,Hinczewski2016} as well as those expected for random-sequence biopolymers \cite{Ferreiro2014}, but occur for a different reason: Heterogeneity and disorder in the IDR occur due to the varying dynamics of structure formation of independent segments, and not because of the topological/energetic frustration effects that dominate the dynamics of certain globular proteins\cite{Frauenfelder1991,Brujic2006,Hinczewski2016} and random-sequence chains \cite{Ferreiro2014}. 

However, there is at least one reported mechanism for nonexponential dynamics in a globular protein that is a more apt comparison: lysozyme was shown to exhibit heterogeneous nucleation dynamics \cite{Morozova-Roche1999}, where different small segments independently form local tertiary structures with varying dynamics; this is analogous to the `foldons' proposed theoretically \cite{Panchenko1996,Klimov1998}. This behavior is similar to that proposed here, with the major difference being the type of structure formed, as tertiary interactions are lacking in the IDR. 
 	
Finally, it has been suggested that many IDRs contain multiple subsegments that form local structures or interactions in an independent, noncooperative fashion \cite{Solanki2014,Uversky2019}. Our analysis indicates that the existence of such distributed, heterogeneous structures is the key feature underlying the slow, logarithmic dynamics. Thus, it is reasonable to expect that glassy relaxations and memory effects could occur broadly in other systems with IDRs.
	
\begin{acknowledgments}
This work was supported by the National Science Foundation under Grant No. 1715627, the Israel Science Foundation under Grant No. 550/15 and 453/17, and the United States-Israel Binational Science Foundation under Grant No. 2016696. We thank Yoav Lahini, Yakov Kantor, and Joan-Emma Shea for useful discussions. 
\end{acknowledgments}
\bibliographystyle{apsrev4-2}
\end{document}